# A quantitative assessment of epidemiological parameters to model COVID-19 burden

Agnese Zardini[1,2,#], Margherita Galli[1,3,#], Marcello Tirani[4], Danilo Cereda[4], Mattia Manica[1], Filippo Trentini[1], Giorgio Guzzetta[1], Valentina Marziano[1], Raffaella Piccarreta[5,6], Alessia Melegaro[5,7], Marco Ajelli[8,9,§], Piero Poletti[1,§,*], Stefano Merler[1,§]

[#] equally contributed
[§] joint senior authors
[*] corresponding author: poletti@fbk.eu

[1] Bruno Kessler Foundation, Trento, Italy
[2] Department of Mathematics, University of Trento, Trento, Italy
[3] Department of Mathematics, Computer Science and Physics, University of Udine, Udine, Italy
[4] Directorate General for Health, Lombardy Region, Milan, Italy
[5] Dondena Centre for Research on Social Dynamics and Public Policy, Bocconi University, Milan, Italy
[6] Department of Decision Sciences, Bocconi University, Milan, Italy
[7] Department of Social and Political Sciences, Bocconi University, Milan, Italy
[8] Department of Epidemiology and Biostatistics, Indiana University School of Public Health, Bloomington, IN, USA
[9] Laboratory for the Modeling of Biological and Socio-technical Systems, Northeastern University, Boston, MA, USA

## Abstract

Solid estimates describing the clinical course of SARS-CoV-2 infections are still lacking due to under-ascertainment of asymptomatic and mild-disease cases. In this work, we quantify age-specific probabilities of transitions between stages defining the natural history of SARS-CoV-2 infection from 1,965 SARS-CoV-2 positive individuals identified in Italy between March and April 2020 among contacts of confirmed cases. Infected contacts of cases were confirmed via RT-PCR tests as part of contact tracing activities or retrospectively via IgG serological tests and followed-up for symptoms and clinical outcomes. In addition, we provide estimates of time intervals between key events defining the clinical progression of cases as obtained from a larger sample, consisting of 95,371 infections ascertained between February and July 2020. We found that being older than 60 years of age was associated with a 39.9% (95%CI: 36.2-43.6%) likelihood of developing respiratory symptoms or fever ≥37.5 °C after SARS-CoV-2 infection; the 22.3% (95%CI: 19.3-25.6%) of the infections in this age group required hospital care and the 1% (95%CI: 0.4-2.1%) were admitted to an intensive care unit (ICU). The corresponding proportions in individuals younger than 60 years were estimated at 27.9% (95%CI: 25.4-30.4%), 8.8% (95%CI: 7.3-10.5%) and 0.4% (95%CI: 0.1-0.9%), respectively. The infection fatality ratio (IFR) ranged from 0.2% (95%CI: 0.0-0.6%) in individuals younger than 60 years to 12.3% (95%CI: 6.9-19.7%) for those aged 80 years or more; the case fatality ratio (CFR) in these two age classes was 0.6% (95%CI: 0.1-2%) and 19.2% (95% CI: 10.9-30.1%), respectively. The median length of stay in hospital was 10 (IQR 3-21) days; the length of stay in ICU was 11 (IQR 6-19) days. The obtained estimates provide insights into the epidemiology of COVID-19 and could be instrumental to refine mathematical modeling work supporting public health decisions.

## Author summary

Mathematical modeling has been extensively adopted to support public decision making in the response to the coronavirus disease pandemic. Metrics describing the potential clinical progression after the infection from the highly transmissible severe acute respiratory syndrome coronavirus 2 are key to properly quantify and interpret the disease burden. We provide a comprehensive quantitative assessment of all the main epidemiological parameters essential to model the natural history of the disease and evaluate the consequent pressure on healthcare systems. To do this, we analyzed epidemiological records of all the laboratory confirmed infections identified between February and July 2020 in Lombardy, Italy. Age and sex specific risk outcomes (e.g., deaths, severe disease, respiratory symptoms) upon infection were estimated on the basis of a highly detailed sample of infections that does not suffer the usual limitations of surveillance data (i.e., underestimation of asymptomatic individuals) and of serological data (i.e., lack of longitudinal records about the clinical history of each individual). The sample consists of positive individuals among contacts of confirmed infections, who were tested via nasal-swab or serological assays irrespective of their symptoms and monitored for their clinical course after exposure to the infection. Time intervals describing the clinical progression of cases, such as the length of stay in hospital and intensive care units, were estimated by analyzing the entire line list of infections ascertained in the region. Our estimates can be instrumental to refine model estimates focusing on the impact of vaccination efforts in the forthcoming months and assisting the design of appropriate strategies to control the pandemic until a sufficiently large proportion of the population become immune.

# Introduction

Mathematical modeling has been one of the cornerstones in the response to the COVID-19 pandemic [1-11]. To provide solid estimates, models need to be properly calibrated based on empirical evidence [7,12-15]. While a lot of work has been done in this direction [14-21], metrics required to estimate the disease burden are still poorly quantified [22,23]. Difficulties in deriving these quantities are related to challenges in defining unbiased denominators (i.e., the infections) for computing different risk outcomes (e.g., deaths, severe disease, respiratory symptoms) upon infection [24-26]. Indeed, as asymptomatic cases and patients with mild disease are, in general, more likely to remain undetected, quantitative estimates of the clinical course of the infection based only on confirmed cases could result in risk outcomes biased upward [12,22,24-26].

In this work, we provide estimates of the probabilities of transition across the stages characterizing the clinical progression after SARS-CoV-2 infection, stratified by age and sex, as well as of the time delays between key events. To do this, we analyzed a sample of 1,965 SARS-CoV-2 positive individuals who were contacts of confirmed cases. These individuals were identified irrespective of their symptoms as part of contact tracing activities carried out in Italy over the period from March 10 to April 27, 2020. These individuals were daily monitored for symptoms for at least two weeks after exposure to a COVID-19 case and either tested for SARS-CoV-2 via PCR or IgG serological assays; their clinical history was also recorded. In addition to this highly detailed sample, we relied on the epidemiological records of 95,371 SARS-CoV-2 PCR confirmed infections reported to the surveillance system between February and July 2020 to provide a comprehensive quantitative assessment of all the main epidemiological parameters essential to model COVID-19 burden, thus laying the foundation for future COVID-19 modeling efforts.

# Results

### Sample description
We analyzed a total of 95,371 laboratory confirmed infections. Of these, 88,538 (92.8%, median age 65 years, IQR: 50 -81) reported respiratory symptoms (details in the Methods) or fever ≥37.5 °C, 47,393 (49.7%, median age 69 years, IQR: 55-80) were hospitalized, 19,020 (19.9%, median age 79 years, IQR: 70-86) developed critical disease (i.e., requiring ICU treatment or resulting in a fatal outcome) and 16,778 (17.6%, median age 81 years, IQR: 73-87) died with a diagnosis of SARS-CoV-2 (Table 1). 3,855 out of 47,393 inpatients had inconsistent data entries on their temporal clinical progression after hospital admission and were therefore excluded when estimating time to key changes in inpatients' status, such as hospital or ICU admission and discharge.

By combining the regional line list of all ascertained infections with contact-tracing records, we obtained a subsample of 1,965 (median age 53 years, IQR: 32-64) contacts who resulted positive to SARS-CoV-2. None of these records showed inconsistent data entries. Among 1,965 infected contacts, 630 (32.1%, median age 57 years, IQR: 42.5 - 71) developed symptoms, 266 (13.5%, median age 64 years, IQR: 53.25 - 76) were hospitalized, 43 (2.2%, median age 76 years, IQR: 69 - 81) experienced critical disease conditions, 12 (0.6%, median age 68 years, IQR: 52.5 -72) were admitted to ICUs, and 35 (1.8%, median age 78 years, IQR: 74.5 – 82.5) resulted in a fatal outcome; 31 (1.6%, median age 79 years, IQR: 75-84) subjects died without being admitted to ICU; 4 (0.2%, median age 73.5 years, IQR: 71.25-75) died after an ICU admission (Table 2). Figure 1 provides a schematic representation of all metrics considered to quantify COVID-19 burden.

### Metrics of COVID-19 burden

Age-specific transition probabilities characterizing the different outcomes after SARS-CoV-2 infection were estimated by considering infections ascertained among close case contacts (details in the Methods). We found that the likelihood of developing respiratory symptoms or fever ≥37.5 °C after SARS-CoV-2 infection (SR) was 27.9% (95%CI: 25.4-30.4%) under 60 years of age and 39.9% (95%CI: 36.2-43.6%) above (see Table 2). We estimated that, in the first age-group, 8.8% (95%CI: 7.3-10.5%) of infected individuals required hospital care (HR) and 0.4% (95%CI: 0.1-0.9%) were admitted to ICU (IR); the corresponding proportions in positive individuals older than 60 years were 22.3% (95%CI: 19.3-25.6%) and 1% (95%CI: 0.4-2.1%), respectively. A significantly higher risk of developing critical disease after infection (CR) was found above 60 years of age when compared to younger individuals: 5.3% (95%CI: 3.7-7.2%) vs 0.5% (95%CI: 0.2-1.1%). The infection fatality ratio (IFR) ranged between 0.2% (95%CI: 0.0-0.6%) in subjects younger than 60 years to 12.3% (95%CI: 6.9-19.7%) for those aged 80 years or more. The case fatality ratio (CFR) in these two age groups was 0.6% (95%CI: 0.1-2%) and 19.2% (95% CI: 10.9-30.1%). Although the case fatality ratio was higher for subjects older than 80 years compared to cases aged 60-79 years (namely, 9.5%, 95%CI: 5.8-14.4%), a significantly lower proportion of ICU admissions was found for the oldest age segment: 1.2% (95%CI: 0.5-2.5%) vs 0% (95%CI: 0-3.2%). A detailed age-stratification of all these quantities is provided in Table 2. The strong age dependency in the risk of developing symptoms and most severe outcomes after SARS-CoV-2 infection was confirmed by a statistical analysis based on generalized linear models applied to infected case contacts (1,965 subjects) and accounting for possible confounding factors (see Table S1 and Methods). The regression analysis also highlighted a significantly higher risk ratio (RR) of hospital admission (RR: 1.34, 95%CI: 1.07-1.67), critical disease (RR: 2.16, 95%CI: 1.17-3.98), and death (RR: 2.15, 95%CI: 1.08-4.27) for infected males as compared to females (Table S1).

### Temporal changes in the investigated risk metrics

Temporal changes in the risk of being admitted to hospital and ICUs were explored by analyzing records of the complete list of 88,538 symptomatic cases, thus including inpatients with inconsistencies in dates defining the temporal clinical progression after hospitalization. Crude ratios computed from ascertained symptomatic cases should be carefully interpreted because of possible biases due to higher ascertainment rates among more severe cases. However, the analysis of this large sample highlighted an increase of admission rates at different levels of intensity of care among the elderly (Figure 2). In particular, hospital admission ratios among ascertained symptomatic cases (asHR) aged more than 80 years increased from the 26.4% (95%CI: 25.5-27.2%) observed between April and May to 34.7% (95%CI: 30.5-39.1%) afterwards. Similarly, the ICU admission ratio among patients hospitalized (hIR) in this age group raised from the 0.9% (95%CI: 0.7-1.1%) observed between March and April to 2.3% (95%CI: 1.2-3.8%) afterwards.

### Time to key events

Time delays from symptom onset to diagnosis and death were investigated by analyzing all the 88,538 symptomatic infections ascertained in Lombardy between February and July 2020 (see Table 1, and Tables S2 for sample description). The temporal clinical progression of inpatients was investigated by analyzing 43,538 hospitalized cases (Table 3), after having excluded 3,855 out of the 47,393 available inpatient records because of inconsistent dates of hospital or ICU admission/discharge (details in the Methods). We estimated that, overall, the median delay between symptom onset and diagnosis was 4 (IQR 1-10) days. The median time from symptoms to death was 12 (IQR 7-21) days. Hospitalization of cases occurred 5 (IQR 2-9) days after patients' symptom onset; admission to ICU occurred 10 (IQR 6-15) days after symptom onset. The median time between hospital and ICU admission was 3 (IQR 0-6) days. The median hospital length of stay was 10 (IQR 3-21) days, while the median length of stay in ICU was 11 (IQR 6-19) days. A negative binomial distribution was used to separately fit each time interval of interest (Figure S2).

When looking at these variables across different ages, we found a shorter delay between symptoms and death in individuals older than 70 years (11-12 days vs 15-16 days at younger ages) and a shorter length of stay in ICU among patients aged 80 years or more (5 days vs 9-12 days at younger ages). We separately analyzed these quantities for cases who developed symptoms before and after March 16, 2020, corresponding to the peak in the number of hospitalized patients in Lombardy (Figure S1). We found a marked decrease in the time required to both diagnose and hospitalize COVID-19 patients after this date (from 7 to 2 days and from 7 to 3 days, respectively, Table 3). Detailed estimates obtained on the temporal clinical progression of COVID-19 cases are reported in Table 3.

## Discussion

In this work, we provided a comprehensive assessment of the parameters regulating COVID-19 burden and natural history. The proposed analysis leveraged data on the infections ascertained in Italy between February and July 2020 to estimate the time between key events and the age- and sex- specific stage-to-stage transition probabilities characterizing the clinical progression of COVID-19.

Previous studies have highlighted that a significant share of SARS-CoV-2 infections is represented by symptom-free subjects and by individuals developing mild disease [7,22,25,27,28]. Therefore, using the number of notified or confirmed COVID-19 cases as an approximation of the number of infections would likely lead to overestimate the risk of disease and severe outcomes, undermining the comparability and generalizability of the obtained results. An illustrative example of the huge uncertainty caused by this phenomenon is provided by the high variability around the available estimates of the proportion of symptomatic infections, ranging from 3% to 87% [25,27,29-32]. To reduce potential biases in the identification of SARS-CoV-2 infections, we estimated different risk ratios based on a sample of SARS-CoV-2 positive individuals who were identified as contacts of confirmed cases and tested irrespectively of their symptoms. A larger sample, consisting of all notified symptomatic cases, was used only to estimate the time to key events describing the clinical progression of cases and to highlight temporal changes in the risk of hospitalization and ICU admission.

Our results confirmed findings from other studies on the strong age gradient in the likelihood of developing symptoms, critical disease, and death after infection [7,24,26,33,34]. The estimated proportion of symptomatic cases among SARS-CoV-2 infections is within the range of estimates obtained in previous studies [29,32] and particularly close to findings obtained in Emery et al. [27]. Our estimated CFR was lower compared to the one obtained in a previous study on other Italian data [33], but slightly higher than those observed in other countries [24,34-36]. The estimated age-profile of the IFR closely resembles Verity et al. [24]. However, our aggregate (population-level) estimate of the IFR is generally higher than those obtained in other studies [7,24,37,38]. Such difference can be due to a variety of factors. First, Italy is one of the oldest countries in the world (average age: 45.7 years [39]). Second, there may be between-country differences in the age distribution of SARS-CoV-2 infected individuals. Third, there are differences in the definition of COVID-19 death. In fact, in Italy, deaths occurring among SARS-CoV-2 positive subjects are classified as COVID-19-related deaths regardless of other conditions that might have caused the observed fatal outcome [33]. This has possibly led to overestimate the number of deaths caused by SARS-CoV-2, especially in the oldest segment of the population.

The proportion of hospitalized patients among positive individuals older than 60 years was almost double than that observed in France [7]. On the other hand, the proportion of ICU admissions and deaths among hospitalized cases was significantly lower in our sample (4.5% and 12.4% vs 19% and 18.1%, respectively), and we found a marked temporal decreasing pattern in the risk of hospital admission among ascertained symptomatic cases. This suggests that hospitalization criteria might have been strongly heterogeneous across different countries and may also greatly vary over time.

The estimated time from symptoms onset to laboratory diagnosis well compares with estimates obtained from Belgian patients [40]. Although in line with previous findings from Belgium [40], the time from symptoms to hospitalization we found is markedly shorter than those observed in France, in China and in the US [7,41,42]. This may be the consequence of the higher proportion of severe cases observed in Italy compared to other countries, which strictly relates to the older age-structure characterizing this country. This hypothesis is partially supported by the shorter hospital length of stay and by the longer length of stay in ICU we found, compared to estimates from China [41,43].

The relatively low ICU admission ratio we observed among the elderly was already highlighted in previous studies [7,44]. However, our findings clearly show that hospitalized patients aged 80 years or older faced the highest risk of fatal outcome, but also the lowest likelihood of being admitted to ICU. The increased ratio of ICU admission among inpatients we found for this age group after April 2020 suggests that elective ICU admission has been initially adopted in Lombardy due to saturation of healthcare resources. The lower delay between symptom onset and admission to hospital observed after March 16, 2020, and the progressive temporal increase in the likelihood of hospital admission among older patients strongly suggest that reducing the pressure on the regional healthcare system markedly improved its capacity to rapidly identify and treat severe patients [45] (see Figure S1).

Our estimates of the risk ratios of hospital and ICU admissions after infection have to be interpreted cautiously. In fact, rather than being purely biological features, such quantities strongly depend on the available healthcare resources, on the temporal changes in the number of patients seeking care, and on the protocols adopted to face a brisk upsurge of COVID-19 cases. Additionally, due to temporal changes in the ascertainment rates of infections, we were not able to quantify the reduced risk of severe outcomes determined by timely detection, diagnosis and treatment of cases, nor to evaluate the role played by the progressive enhancement in the treatment procedures in reducing the risk of disease. A further limitation affecting our study is the lack of data to disentangle the role played by patients' comorbidities in shaping the risk of severe diseases. Finally, it is important to stress that estimates reported here are associated with the historical and dominant variant of the virus that circulated during 2020, in the absence of vaccination. As such, estimated metrics may not apply to new emerging SARS-CoV-2 variants [46-48] and may not reflect the risk of developing COVID-19 disease among infections occurring among vaccinated individuals.

Despite the aforementioned limitations, the provided metrics can be instrumental to refine model estimates. In particular, our findings could be used to assist the design and evaluation of forthcoming vaccination efforts and the development of appropriate strategies to control the COVID-19 pandemic until a sufficiently large proportion of the population has become immune.

# Methods

### Study population

Lombardy represents the earliest and most affected region by the first COVID-19 epidemic wave experienced in Italy. Short after the detection of a first COVID-19 case on February 20, 2020, a ban of mass gatherings and the suspension of teaching in schools and universities was applied to the entire region. The interruption of non-essential productive activities and strict individual movement restrictions were imposed to the most affected municipalities. On March 8, 2020, after a rapid increase of cases, closure of all non-necessary businesses and industries and limitations of movements except in cases of necessity were extended to the entire region. A national lockdown was imposed on March 10, 2020. Suspended economic and social activities were gradually resumed between April 14 and May 18, 2020.

### Data collection

The main data analyzed here consists of the line list of all SARS-CoV-2 laboratory confirmed infections ascertained in Lombardy between February 20 and July 16, 2020, and regularly updated by the regional public health authorities. Information retrieved from this dataset was complemented with contact-tracing records collected between March 10 and April 27 and with results of a serological survey targeting case contacts conducted between April 16 to June 15, 2020 [25, 26]. Data collection, integration, storage, and anonymization was managed by regional health authorities as part of surveillance activities and outbreak investigations aimed at controlling and mitigating the COVID-19 epidemic in Italy.

### Definition of COVID-19 case

From February 21 to February 25, following the criteria initially defined by the European Centre for Disease Prevention and Control (ECDC), suspected COVID-19 cases were identified as:
1. patients with acute respiratory tract infection OR sudden onset of at least one of the following: cough, fever, shortness of breath AND with no other aetiology that fully explains the clinical presentation AND at least one of these other conditions: a history of travel to or residence in China, OR patients among health care workers who has been working in an environment where severe acute respiratory infections of unknown etiology are being cared for;
2. OR patients with any acute respiratory illness AND at least one of these other conditions: having been in close contact with a confirmed or probable COVID-19 case in the last 14 days prior to onset of symptoms, OR having visited or worked in a live animal market in Wuhan, Hubei Province, China in the last 14 days prior to onset of symptoms, OR having worked or attended a health care facility in the last 14 days prior to onset of symptoms where patients with hospital-associated COVID-19 have been reported.

Probable cases were defined as suspect cases testing positive with a specific real-time reverse transcription polymerase chain reaction (RT-PCR) assay targeting different genes of SARS-CoV-2 [18,49,50]. From March 20, positivity to the nasopharyngeal swab was also granted for tests that sought a single gene. At any time, ascertained infections were defined as persons with laboratory confirmation of virus causing of SARS-CoV-2 infection, irrespective of clinical signs and symptoms. Inconclusive swabs were repeated to reach the diagnosis.

**Ascertainment of infections among close case contacts**

All ascertained SARS-CoV-2 infections were considered as potential index cases for further spread of SARS-CoV-2. Close contacts of these individuals were therefore identified through standard interviews of cases, informed of their possible exposure and quarantined within 24-48 hours from a positive test result on the index case.

A close case contact was defined as a person living in the same household as a COVID-19 confirmed case; a person having had face-to-face interaction with a COVID-19 confirmed case within 2 meters and for more than 15 minutes; a person who was in a closed environment (e.g. classroom, meeting room, hospital waiting room) with a COVID-19 confirmed case at a distance of less than 2 meters for more than 15 minutes; a healthcare worker or other person providing direct care for a COVID-19 confirmed case, or laboratory workers handling specimens from a COVID-19 confirmed case without recommended personal protective equipment (PPE) or with a possible breach of PPE; a contact in an aircraft sitting within two seats (in any direction) of a COVID-19 confirmed case, travel companions or persons providing care, and crew members serving in the section of the aircraft where the index case was seated (passengers seated in the entire section or all passengers on the aircraft were considered close contacts of a confirmed case when severity of symptoms or movement of the case indicate more extensive exposure). Close case contacts were initially considered as contacts occurred between 14 days before and 14 days after the date of symptom onset of the index case of the cluster. After March 20, the exposure period was shortened, ranging from 2 days before to 14 days after the symptom onset of the index case [51]. For individuals unable to sustain the contact tracing interview, close contacts were identified by their parents, relatives or their emergency contacts. From February 20 to February 25, 2020 all contacts of confirmed infections were tested with RT-PCR, irrespective of clinical symptoms. From February 26 onward, the traced contacts were tested only in case of symptom onset.

On April 16, 2020, regional health authorities initiated an IgG serological survey of quarantined case contacts without history of testing against SARS-CoV-2 infection. The test used to detect SARS-CoV-2 IgG antibodies was the LIAISONR SARS-CoV-2 test (DiaSorin), employing magnetic beads coated with S1 & S2 antigens. The antigens used in the tests are expressed in human cells to achieve proper folding, oligomer formation, and glycosylation, providing material similar to the native spikes. The S1 and S2 proteins are both targets to neutralizing antibodies. The test provides the detection of neutralizing antibodies with 94.4% sensitivity at 15 days from diagnosis and 98.3% specificity. Performance analyses validating the accuracy of this serological test can be found in [52]. Serological test results were binary and communicated to tested participants, who were categorized as seropositive if they had developed IgG antibodies.

All case contacts, irrespectively to the presence of a laboratory diagnosis, were followed up for at least 14 days after exposure to an index case and required by national regulations to report symptoms to local public health authorities. Symptomatic cases were defined as infected subjects showing fever ≥37.5 °C or one of the following symptoms: dry cough, dyspnea, tachypnea, difficulty breathing, shortness of breath, sore throat, and chest pain or pressure. The definition of symptoms did not change throughout the period considered in this study. Clinical manifestations, admission to hospital or intensive care units and death among both ascertained infections and their close contacts were regularly updated by the regional health surveillance. In our study, critical cases were defined as patients either admitted to an intensive care unit or had a fatal outcome with a diagnosis of SARS-CoV-2 infection.

### Sample selection for computing risk outcomes

A large fraction of case contacts remained untested against SARS-CoV-2 infection, due to difficulties in maintaining a high level of testing during the contact tracing operations and to the relatively low coverage of IgG serological screening conducted on traced contacts. As asymptomatic infections ascertained by surveillance systems are likely under-represented, we selected a subsample of SARS-CoV-2 positive individuals who were tested irrespectively from their symptoms. In particular, we considered infections ascertained case contacts identified between March 10 and April 27 and belonging to clusters whose individuals were all daily followed up for symptoms and tested either via RT-PCR or IgG serological assays. This procedure allowed us to minimize the risks of bias in the identification of infections when computing the proportion of SARS-CoV-2 infections developing symptoms and severe conditions. The resulting subsample consisted of 1,965 positive subjects identified in 2,458 clusters of 3,947 close contacts.

### Statistical analysis

The aforementioned subsample of 1,965 positive individuals who were identified as contacts of confirmed cases was analyzed to estimate the likelihood of developing respiratory symptoms or fever ≥37.5 °C, of being admitted to a hospital and an ICU, of developing critical disease and of dying after SARS-CoV-2 infection. Age and sex specific ratios were computed as crude percentages; 95% confidence intervals were computed by exact binomial tests. Logistic regression models were used to estimate the corresponding risk ratios (RRs) using the case age group, sex and month of identification (March or April) as model covariates. For the regression analysis, the following age-groups were considered: 0-59 years, 60-74 years, 75+ years.

The entire sample of cases ascertained by regular surveillance activities (88,538 symptomatic individuals) was used to investigate temporal changes in the COVID-19 disease burden. In particular, we computed the age-specific crude percentage of ascertained cases admitted to hospital and the percentage of ICU admissions among hospitalized cases for four epidemic periods: before April, April, May and after May.

The same sample of cases was used to investigate the distribution of patients' length of stay in hospital and in ICU, and the time interval between the following key events: from symptoms onset to diagnosis, from symptoms to hospital and ICU admission, from symptoms to death, and from hospitalization to ICU admission. As 3,855 out of 47,393 inpatients had inconsistent data entries on their temporal clinical progression after hospital admission, we excluded the corresponding data records when estimating time to key changes in patients' status, such as hospital or ICU admission and discharge. Specifically, we excluded inpatients with a date of hospital admission or of death preceding the date of symptom onset, patients with a date of ICU admission or death preceding their hospitalization and patients with a negative length of stay in ICU or in hospital. Estimates for the hospital and ICU length of stay and the time between key events are provided for two epidemic periods, defined by considering the date of peak in the COVID-19 incidence experienced during the first epidemic wave in Lombardy, namely March 16, 2020. Cases were aggregated on the basis of the initial date of the considered interval. Negative binomial distributions were used to separately fit each time interval of interest.

The statistical analysis was performed with the software R (version 3.6), using the "MASS" package.

### Ethical statement

Data collection and analysis were part of outbreak investigations during a public health emergency. Processing of COVID-19 data is necessary for reasons of public interest in the area of public health, such as protecting against serious cross-border threats to health or ensuring high standards of quality and safety of health care, and therefore exempted from institutional review board approval (Regulation EU 2016/679 GDPR).

### Author Contributions

PP, MA, SM conceived and designed the study. AZ and MG performed the analysis. MT and DC collected data. AZ, MG, MT, DC, FT, RP and PP collated linked epidemiological data. MT, DC verified all data. AZ, MG and PP wrote the first draft. All authors contributed to data interpretation, critical revision of the manuscript and approved the final version of the manuscript.

### Competing Interest

MA has received research funding from Seqirus. The funding is not related to COVID-19. All other authors declare no competing interest.

### Funding

MT, FT, VM, GG, PP and SM acknowledge funding from EU grant 874850 MOOD (catalogued as MOOD 000). The funders had no role in study design, data collection and analysis, decision to publish, or preparation of the manuscript.

# Figures and Tables

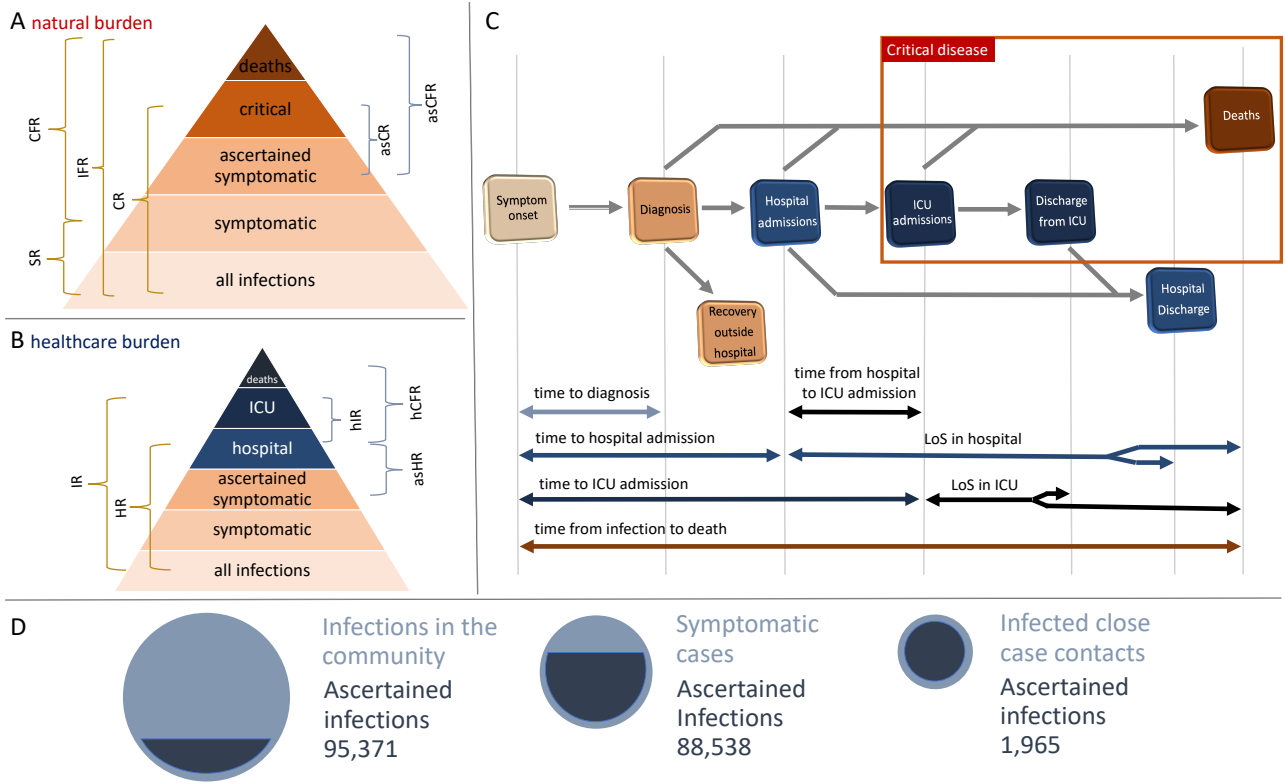

**Figure 1.** **A** Schematic representation of transition probabilities characterizing possible disease outcomes after SARS-CoV-2 infection. These include the symptomatic ratio (SR), the ratio of critical cases (CR), the case (CFR) and infection fatality ratios (IFR) and similar quantities that could be estimated using ascertained symptomatic infections (asCR, asCFR) as the set of exposed individuals. **B** Schematic representation of transition probabilities characterizing the hospital (HR) and ICU (IR) admission among infected individuals, and of similar quantities that could be estimated using ascertained symptomatic infections (asHR) or hospital patients (hCFR, hIR) as the set of exposed individuals. **C** Schematic representation of time to key events defining the temporal clinical progression of cases. **D** Schematic representation of the marked differences expected in the ascertainment rates associated with SARS-CoV-2 infections occurred in the community, infections who developed COVID-19 symptoms and close case contacts who were all tested for SARS-CoV-2 infection and daily monitored for symptoms during their quarantine or isolation period.

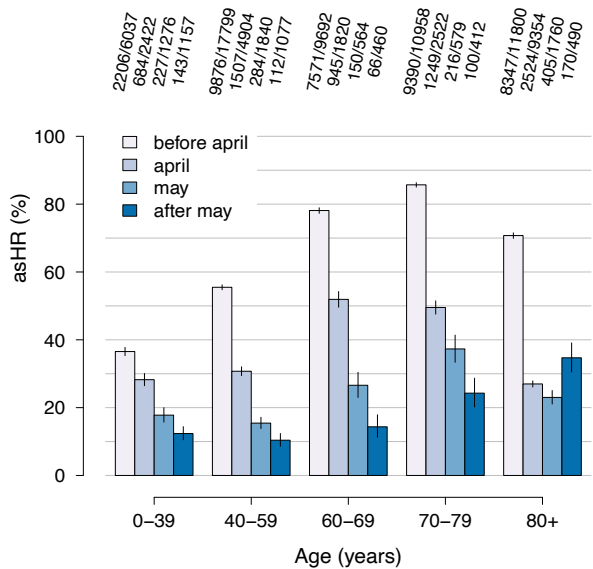 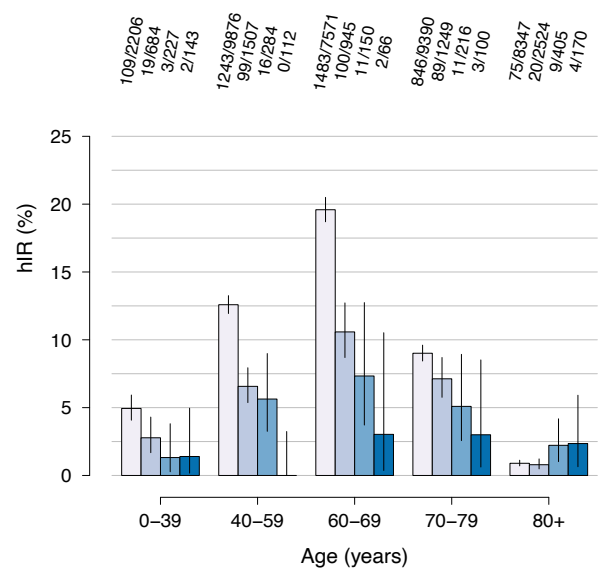

Figure 2. **A** Age-specific case hospital admission ratios among ascertained symptomatic cases (asHR). **B** Age-specific ICU admission ratios among hospitalized cases (hIR). Bars of different colors represent crude percentages observed across different epidemic periods; vertical lines represent 95% confidence intervals computed by exact binomial tests. Numbers shown in each panel represent the age-specific number of events observed in the data among exposed COVID-19 cases.

Table 1. Estimated risk ratios of hospital admission, experiencing critical disease, and fatal outcome among symptomatic cases, disaggregated by age, sex, and period.

|  | Symptomatic cases | Hospitalized patients | | Critical cases | | Deaths | |
|---|---|---|---|---|---|---|---|
|  | Count | Count | Risk ratio (95%CI) | Count | Risk ratio (95%CI) | Count | Risk ratio (95%CI) |
| **Age** | | | | | | | |
| ≥ 80 | 24,092 | 11,849 | Reference | 9,325 | Reference | 9,291 | Reference |
| 0-39 | 11,019 | 3,361 | 0.54 (0.52 - 0.56) | 158 | 0.03 (0.03 - 0.04) | 38 | 0.01 (0.01 - 0.01) |
| 40-59 | 25,910 | 12,037 | 0.77 (0.75 - 0.79) | 1,747 | 0.12 (0.12 - 0.13) | 733 | 0.05 (0.05 - 0.06) |
| 60-69 | 12,731 | 8,917 | 1.22 (1.19 - 1.24) | 2,642 | 0.35 (0.33 - 0.37) | 1,872 | 0.24 (0.23 - 0.26) |
| 70-79 | 14,784 | 11,228 | 1.39 (1.37 - 1.41) | 5,147 | 0.65 (0.63 - 0.68) | 4,844 | 0.62 (0.60 - 0.64) |
| Unknown | 2 | 1 | 1.13 (0.06 - 1.99) | 1 | 0.86 (0.05 - 2.40) | 0* | - |
| **Sex** | | | | | | | |
| Female | 46,234 | 19,318 | Reference | 7,323 | Reference | 6,804 | Reference |
| Male | 42,168 | 28,061 | 1.49 (1.47 - 1.51) | 11,682 | 1.87 (1.82 - 1.92) | 9,966 | 1.78 (1.73 - 1.84) |
| Unknown | 136 | 14 | 0.20 (0.11 - 0.33) | 15 | 1.28 (0.75 - 1.97) | 8 | 0.84 (0.38 - 1.57) |
| **Epidemic period** | | | | | | | |
| Before April | 56,288 | 37,391 | Reference | 14,473 | Reference | 12,530 | Reference |
| April | 21,022 | 6,909 | 0.52 (0.51 - 0.54) | 3,451 | 0.49 (0.47 - 0.51) | 3,239 | 0.48 (0.46 - 0.50) |

| | | | | | | | | | |
|---|---|---|---|---|---|---|---|---|---|
| May | 6,019 | 1282 | 0.36 (0.34 - 0.38) | 313 | 0.20 (0.18 - 0.22) | 276 | 0.19 (0.16 - 0.21) | | |
| After May | 3,596 | 591 | 0.27 (0.25 - 0.29) | 44 | 0.06 (0.04 - 0.08) | 39 | 0.06 (0.04 - 0.08) | | |
| Unknown | 1,613 | 1,220 | 1.15 (1.11 - 1.18) | 739 | 1.56 (1.45 - 1.66) | 694 | 1.63 (1.51 - 1.76) | | |

* RR and 95%CI were not computed for insufficiently large sample size

Table 2. Estimated crude percentages of symptomatic, hospitalized, ICU admitted, and critical cases among SARS-CoV-2 positive individuals who were identified as contacts of confirmed cases as well as estimated risk of death among positive individuals (i.e., infections) and symptomatic case (i.e., infected and symptomatic) individuals who were identified as contacts of confirmed cases. Results are disaggregated by age and sex.

| | Positives contacts | Symptomatic cases | | Critical cases | | Deaths | | | Hospitalized patients | | ICU-admitted patients | |
|---|---|---|---|---|---|---|---|---|---|---|---|---|
| | Count | Count | Proportion (95% CI) | Count | Proportion (95% CI) | Count | IFR (95% CI) | CFR (95% CI) | Count | Proportion (95% CI) | Count | Proportion (95% CI) |
| **Age** | | | | | | | | | | | | |
| 0-14 | 219 | 39 | 17.8% (13-23.5%) | 0 | 0% (0-1.7%) | 0 | 0% (0-1.7%) | 0% (0-9%) | 4 | 1.8% (0.5-4.6%) | 0 | 0% (0-1.7%) |
| 15-19 | 22 | 6 | 27.3% (10.7-50.2%) | 0 | 0% (0-15.4%) | 0 | 0% (0-15.4%) | 0% (0-45.9%) | 2 | 9.1% (1.1-29.2%) | 0 | 0% (0-15.4%) |
| 20-39 | 377 | 99 | 26.3% (21.9-31%) | 2 | 0.5% (0.1-1.9%) | 0 | 0% (0-1%) | 0% (0-3.7%) | 18 | 4.8% (2.9-7.4%) | 2 | 0.5% (0.1-1.9%) |
| 40-59 | 662 | 213 | 32.2% (28.6-35.9%) | 5 | 0.8% (0.2-1.8%) | 2 | 0.3% (0-1.1%) | 0.9% (0.1-3.4%) | 89 | 13.4% (10.9-16.3%) | 3 | 0.5% (0.1-1.3%) |
| 60-69 | 331 | 106 | 32% (27-37.3%) | 5 | 1.5% (0.5-3.5%) | 3 | 0.9% (0.2-2.6%) | 2.8% (0.6-8%) | 49 | 14.8% (11.2-19.1%) | 3 | 0.9% (0.2-2.6%) |
| 70-79 | 240 | 94 | 39.2% (33-45.7%) | 17 | 7.1% (4.2-11.1%) | 16 | 6.7% (3.9-10.6%) | 17% (10.1-26.2%) | 59 | 24.6% (19.3-30.5%) | 4 | 1.7% (0.5-4.2%) |
| ≥ 80 | 114 | 73 | 64% (54.5-72.8%) | 14 | 12.3% (6.9-19.7%) | 14 | 12.3% (6.9-19.7%) | 19.2% (10.9-30.1%) | 45 | 39.5% (30.4-49.1%) | 0 | 0% (0-3.2%) |
| **Sex** | | | | | | | | | | | | |
| Female | 1,111 | 365 | 32.9% (30.1-35.7%) | 19 | 1.7% (1-2.7%) | 16 | 1.4% (0.8-2.3%) | 4.4% (2.5-7%) | 139 | 12.5% (10.6-14.6%) | 5 | 0.5% (0.1-1%) |
| Male | 854 | 265 | 31% (27.9-34.3%) | 24 | 2.8% (1.8-4.2%) | 19 | 2.2% (1.3-3.5%) | 7.2% (4.4-11%) | 127 | 14.9% (12.6-17.4%) | 7 | 0.8% (0.3-1.7%) |

Table 3. Time intervals between key events as estimated from laboratory confirmed infections ascertained in Lombardy between February 20 and July 16, 2020.

| | Median days (IQR) | | | | | | |
|---|---|---|---|---|---|---|---|
| | Time between symptom onset and diagnosis of SARS-CoV-2 | Time between symptom onset and death | Time between symptom onset and hospital admission | Time between symptom onset and ICU admission | Time between hospital and ICU admission | Hospital length of stay | ICU length of stay for |
| **Age** | | | | | | | |
| 0-39 | 3 (0-11) | 16 (6-27) | 4 (1-8) | 8 (4-11) | 1 (0-5) | 4 (0-10) | 9 (4-15.75) |
| 40-59 | 5 (1-11) | 15 (9-26) | 6 (2-10) | 10 (6-13) | 2 (0-6) | 9 (1-18) | 11 (6-19) |
| 60-69 | 5 (1-10) | 16 (9-25) | 6 (2-10) | 11 (7-15) | 3 (0-7) | 13 (6-24) | 12 (6-20) |
| 70-79 | 5 (1-9) | 12 (7-20) | 5 (2-9) | 10 (6-16) | 4 (1-7) | 12 (5-24) | 10 (5-18) |
| ≥ 80 | 3 (0-8) | 11 (6-19) | 4 (1-8) | 9 (4-23) | 2 (0-11.5) | 10 (4-24) | 5 (3-10.75) |
| Unknown | 13 (9.5-16.5) | 0 (0-0) | 8 (8-8) | 8 (8-8) | 0 (0-0) | 14 (14-14) | 14 (14-14) |
| **Epidemic period** | | | | | | | |
| Before 16 March | 7 (3-12) | 13 (8-21) | 7 (3-10) | 10 (7-15) | 3 (0-6) | 10 (3-21) | 11 (6-19) |
| After 16 March | 2 (0-7) | 11 (6-20) | 3 (1-7) | 9.5 (5-14) | 4 (0-8) | 10 (3-23) | 11 (6-20) |
| Unknown | 0 (0-0) | 0 (0-0) | 0 (0-0) | 0 (0-0) | 0 (0-4) | 8 (2-20) | 6 (3.75-13) |
| Overall | 4 (1-10) | 12 (7-21) | 5 (2-9) | 10 (6-15) | 3 (0-6) | 10 (3-21) | 11 (6-19) |
| **Estimates obtained by fitting a negative binomial distribution to observed data** | | | | | | | |
| overdispersion | 0.445 | 1.638 | 0.831 | 2.036 | 0.57 | 0.746 | 1.517 |
| mean | 9.753 | 16.105 | 7.385 | 12.159 | 5.309 | 15.54 | 14.614 |

# Supplementary Figures and Tables

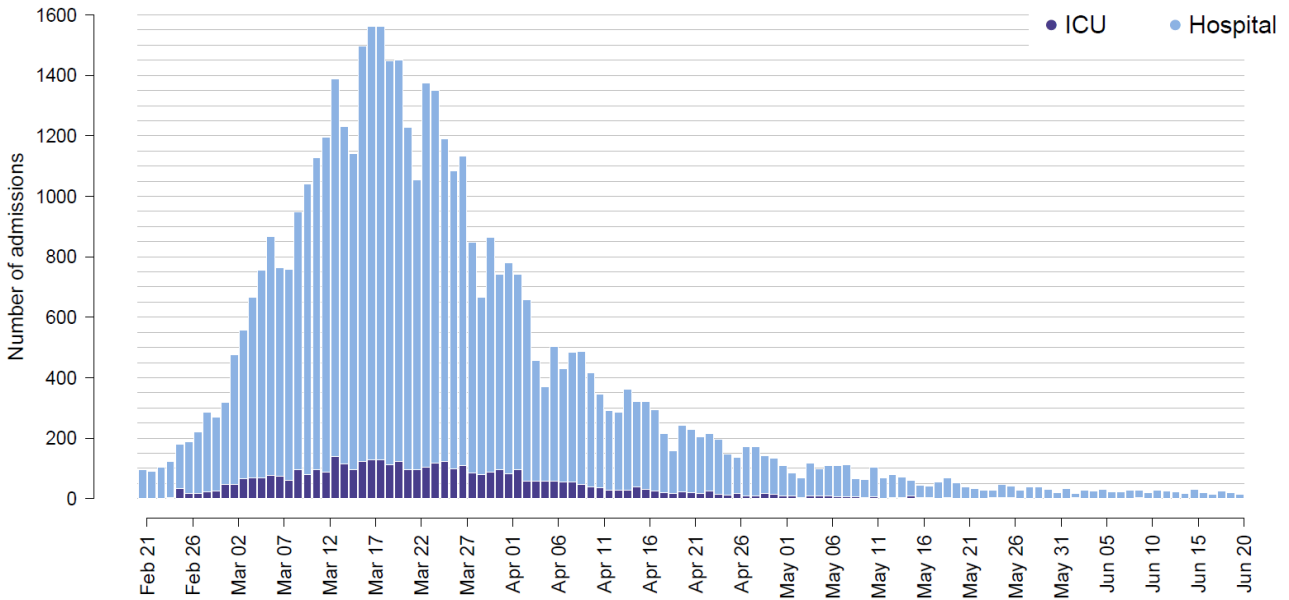

**Figure S1.** Number of hospital (light blue) and ICU (dark blue) admissions of COVID-19 cases in Lombardy, Italy between February 21 and June 20, 2020.

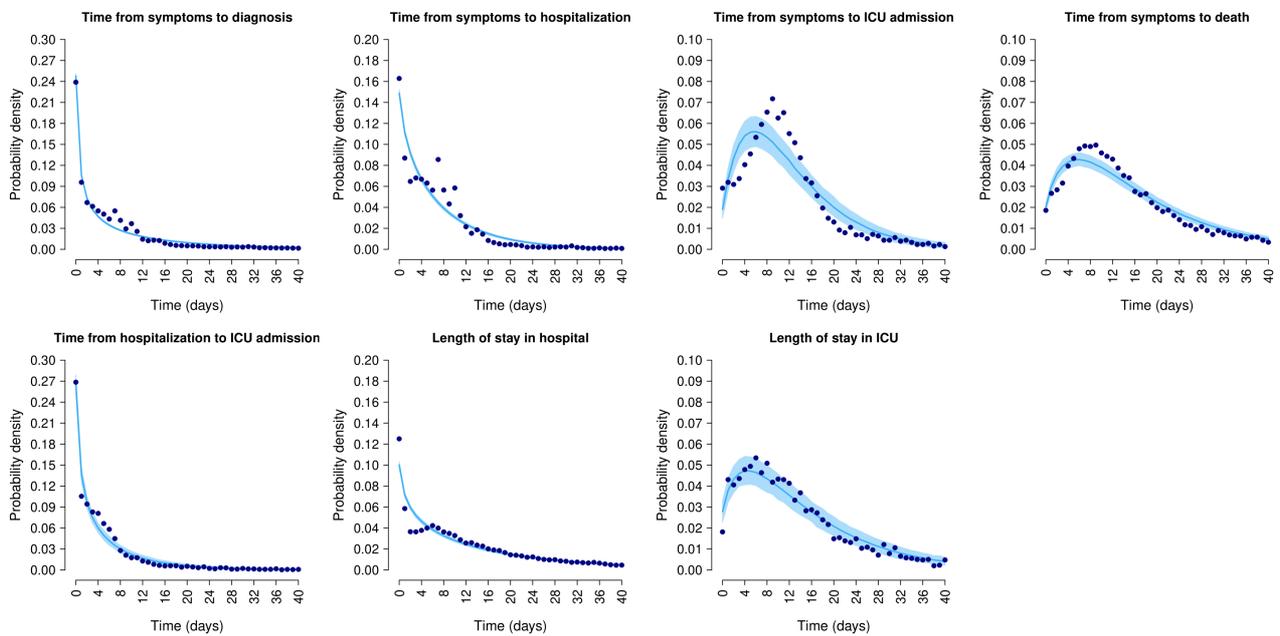

**Figure S2.** Estimated and observed distributions of time intervals between key events. (Top row) Blue dots represent the observed distribution of time from symptoms to diagnosis, hospitalization, ICU admission and death. Light blue lines show the mean frequencies obtained by simulating 1,000 different datasets with size equal to the number of observations in the data on the basis of a negative binomial model. Shaded areas represent the corresponding 95% prediction interval. (Bottom row) As for the top row, but for the time between hospitalization and ICU admission, for the length of stay in hospital and ICU.

Table S1. Estimated risk ratios of developing symptoms, hospital and ICU admission, experiencing critical disease, and fatal outcome among positive contacts of confirmed cases as identified during contact tracing operations. Results are disaggregated by age, sex, and period.

|  | Positives contacts | Symptomatic cases | | Critical cases | | Deaths | | Hospitalized patients | | ICU-admitted patients | |
|---|---|---|---|---|---|---|---|---|---|---|---|
|  | Count | Count | Risk ratio (95%CI) | Count | Risk ratio (95%CI) | Count | Risk ratio (95%CI) | Count | Risk ratio (95%CI) | Count | Risk ratio (95%CI) |
| **Age** | | | | | | | | | | | |
| ≥ 75 | 215 | 118 | Reference | 26 | Reference | 26 | Reference | 75 | Reference | 2 | Reference |
| 0-59 | 1,280 | 357 | 0.51 (0.41-0.62) | 7 | 0.04 (0.02-0.09) | 2 | 0.01 (0.00-0.04) | 113 | 0.24 (0.18-0.33) | 5 | 0.39 (0.08-2.67) |
| 60-74 | 470 | 155 | 0.60 (0.48-0.74) | 10 | 0.18 (0.08-0.36) | 7 | 0.12 (0.05-0.27) | 78 | 0.48 (0.35-0.65) | 5 | 1.19 (0.26-7.86) |
| **Sex** | | | | | | | | | | | |
| Female | 1,111 | 365 | Reference | 19 | Reference | 16 | Reference | 139 | Reference | 5 | Reference |
| Male | 854 | 265 | 0.97 (0.85-1.11) | 24 | 2.16 (1.17-3.98) | 19 | 2.15 (1.08-4.27) | 127 | 1.34 (1.07-1.67) | 7 | 2.04 (0.64-6.86) |
| **Epidemic period** | | | | | | | | | | | |
| March | 1,500 | 467 | Reference | 35 | Reference | 30 | Reference | 200 | Reference | 8 | Reference |
| April | 465 | 163 | 1.14 (0.98-1.31) | 8 | 0.71 (0.30-1.47) | 5 | 0.50 (0.17-1.22) | 66 | 1.07 (0.81-1.38) | 4 | 1.60 (0.43-5.04) |

Table S2. Estimated risk ratios of ICU admission and fatal outcome among hospitalized patients, disaggregated by age, sex, and period

|  | Hospitalized patients | ICU-admitted patients | | Deaths | |
|---|---|---|---|---|---|
|  | Count | Count | Risk ratio (95%CI) | Count | Risk ratio (95%CI) |
| **Age** | | | | | |
| ≥ 80 | 11,849 | 110 | Reference | 6,370 | Reference |
| 0-39 | 3,361 | 136 | 4.49 (3.51 - 5.73) | 31 | 0.02 (0.01 - 0.02) |
| 40-59 | 12,037 | 1,397 | 11.33 (9.53 - 13.55) | 681 | 0.09 (0.08 - 0.10) |
| 60-69 | 8,917 | 1,615 | 16.99 (14.43 - 20.07) | 1,740 | 0.30 (0.29 - 0.32) |
| 70-79 | 11,228 | 965 | 8.03 (6.70 - 9.70) | 4,323 | 0.64 (0.62 - 0.66) |
| Unknown | 1 | 1* | - | 0* | - |
| **Sex** | | | | | |
| Female | 19,318 | 891 | Reference | 4,586 | Reference |
| Male | 28,061 | 3,324 | 2.19 (2.04 – 2.35) | 8,557 | 1.46 (1.41 – 1.50) |
| Unknown | 14 | 9 | 14.79 (8.60 – 19.19) | 2 | 2.59 (0.78 – 3.69) |
| **Epidemic period** | | | | | |
| Before April | 37,391 | 3,757 | Reference | 10,886 | Reference |
| April | 6,909 | 327 | 0.63 (0.56 – 0.71) | 1,708 | 0.72 (0.68 – 0.76) |
| **May** | 1,282 | 50 | 0.53 (0.39 – 0.69) | 158 | 0.36 (0.30 – 0.42) |
| **After May** | 591 | 11 | 0.25 (0.13 – 0.44) | 30 | 0.15 (0.10 – 0.21) |
| **Unknown** | 1,220 | 79 | 0.74 (0.58 – 0.92) | 363 | 0.90 (0.81 – 1.00) |

* RR and 95%CI were not computed for insufficiently large sample size